\documentclass[12pt]{article}

\usepackage{dina4pcg}
\usepackage{epsfig}
\usepackage{amssymb}

\def\oo{\'o}
\def\asz{\as(\mz)}
\def\as{\alpha_s}
\def\mz{M_Z}
\def\z0{Z^0}
\def\qq{q\bar q}
\def\etjet{E_T^{\rm jet}}
\def\etajet{\eta^{\rm jet}}
\def\q2{Q^2}
\def\kt{k_T}
\def\g2{GeV$^2$}
\def\etjb{E^{\rm jet}_{T,{\rm B}}}
\def\etajb{\eta^{\rm jet}_{\rm B}}
\def\asmz#1#2#3#4#5#6{\asz = #1\pm #2\ {\rm (stat.)}\ ^{+#4}_{-#3}\ {\rm (exp.)}\ ^{+#6}_{-#5}\ {\rm (th.)}}
\def\etalab{\eta^{\rm jet}_{\rm LAB}}
\def\yc{y_{\rm cut}}
\def\etal{et al.}
\def\colab#1{#1 Collaboration}

\begin{document}

\title{\bf Jet production in deep inelastic {\boldmath ${ep}$} scattering 
at HERA\footnote{Talk given at the ``Ringberg workshop: New trends in
  HERA Physics 2003'', Ringberg Castle, Germany, $28^{\rm th}$
  September - $3^{\rm rd}$ October, 2003.}}

\author{C. Glasman\thanks{Ram\oo n y Cajal Fellow.}\\
Universidad Aut\oo noma de Madrid, Spain}

\date{}

\maketitle

\begin{abstract}
Recent results from jet production in deep inelastic $ep$ scattering at
HERA are reviewed. The values of $\asz$ extracted from a QCD analysis
of the data are presented.
\end{abstract}

\section{Introduction}
Jet production in neutral-current (NC) deep inelastic $ep$ scattering
(DIS) provides a test of perturbative QCD (pQCD) calculations and of
the parametrisations of the proton parton densities (PDFs). Jet cross
sections allow the determination of the fundamental parameter of QCD,
the strong coupling constant $\as$, and help to constrain the parton
densities in the proton.

Up to leading order (LO) in $\as$, jet production in NC DIS
proceeds via the quark-parton model (QPM) ($Vq\rightarrow q$, where
$V=\gamma$ or $\z0$), boson-gluon fusion (BGF) 
($Vg\rightarrow \qq$) and QCD-Compton (QCDC) ($Vq\rightarrow qg$)
processes. The jet production cross section is given in pQCD by the
convolution of the proton PDFs and the subprocess cross section,
$$d\sigma_{\rm jet}=\sum_{a=q,\bar q,g}\int dx\ f_a(x,\mu_F)\ d\hat \sigma_a(x,\as(\mu_R),\mu_R,\mu_F),$$
where $x$ is the fraction of the proton's momentum taken by the
interacting parton, $f_a$ are the proton PDFs, $\mu_F$ is the
factorisation scale, $\hat\sigma_a$ is the subprocess cross section
and $\mu_R$ is the renormalisation scale.

All the data accumulated from HERA and fixed-target experiments have
allowed a good determination of the proton PDFs over a large phase
space. Then, measurements of jet production in neutral current DIS
provide accurate tests of pQCD and a determination of the fundamental
parameter of the theory, $\as$.

At high scales, calculations using the DGLAP evolution equations have
been found to give a good description of the data up to
next-to-leading order (NLO). Therefore, by fitting the data with these
calculations, it is possible to extract accurate values of $\as$ and
the gluon density of the proton. However, for scales of 
$\etjet\sim Q$, where $\etjet$ is the jet transverse energy and $Q$ is
the exchanged photon virtuality, and large values of the jet
pseudorapidity, $\etajet$, large discrepancies between the data
and the NLO calculations have been observed at low $x$. This could
indicate a breakdown of the DGLAP evolution and the onset of BFKL
effects. These discrepancies can also be explained by assigning a
partonic structure to the exchanged virtual photon or a large
contribution of higher order effects at low $\q2$.

\section{Inclusive jet cross sections}
Inclusive jet cross sections have been measured~\cite{inczeus} in the
Breit frame using the $\kt$-cluster algorithm in the longitudinally
invariant mode. The measurements were made in the kinematic region
given by $\q2>125$ \g2\ and $-0.7<\cos\gamma<0.5$, where $\gamma$ is
the angle of the struck quark in the quark-parton model in the HERA
laboratory frame. The cross sections refer to jets of $\etjb>8$ GeV
and $-2<\etajb<1.8$, where $\etjb$ and $\etajb$ are the jet transverse
energy and pseudorapidity, respectively, in the Breit frame.

The use of inclusive jet cross sections in a QCD analysis presents several
advantages: inclusive jet cross sections are infrared insensitive and
better suited to test resummed calculations and the theoretical
uncertainties are smaller than for dijet cross sections.

\begin{figure}[h]
\setlength{\unitlength}{1.0cm}
\begin{picture} (10.0,11.0)
\put (-0.2,-0.5){\epsfig{figure=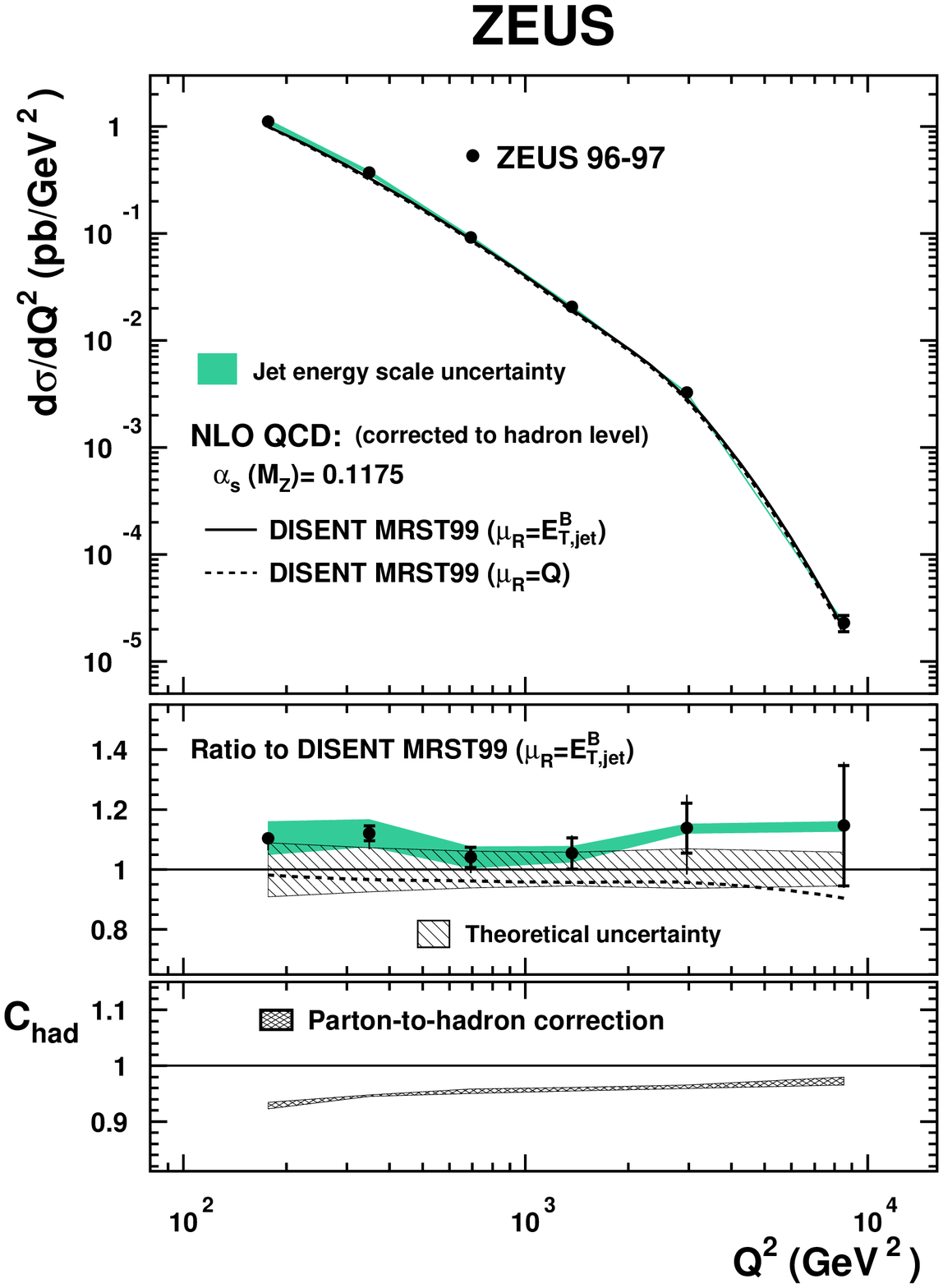,width=9.5cm}}
\put (8.5,-0.5){\epsfig{figure=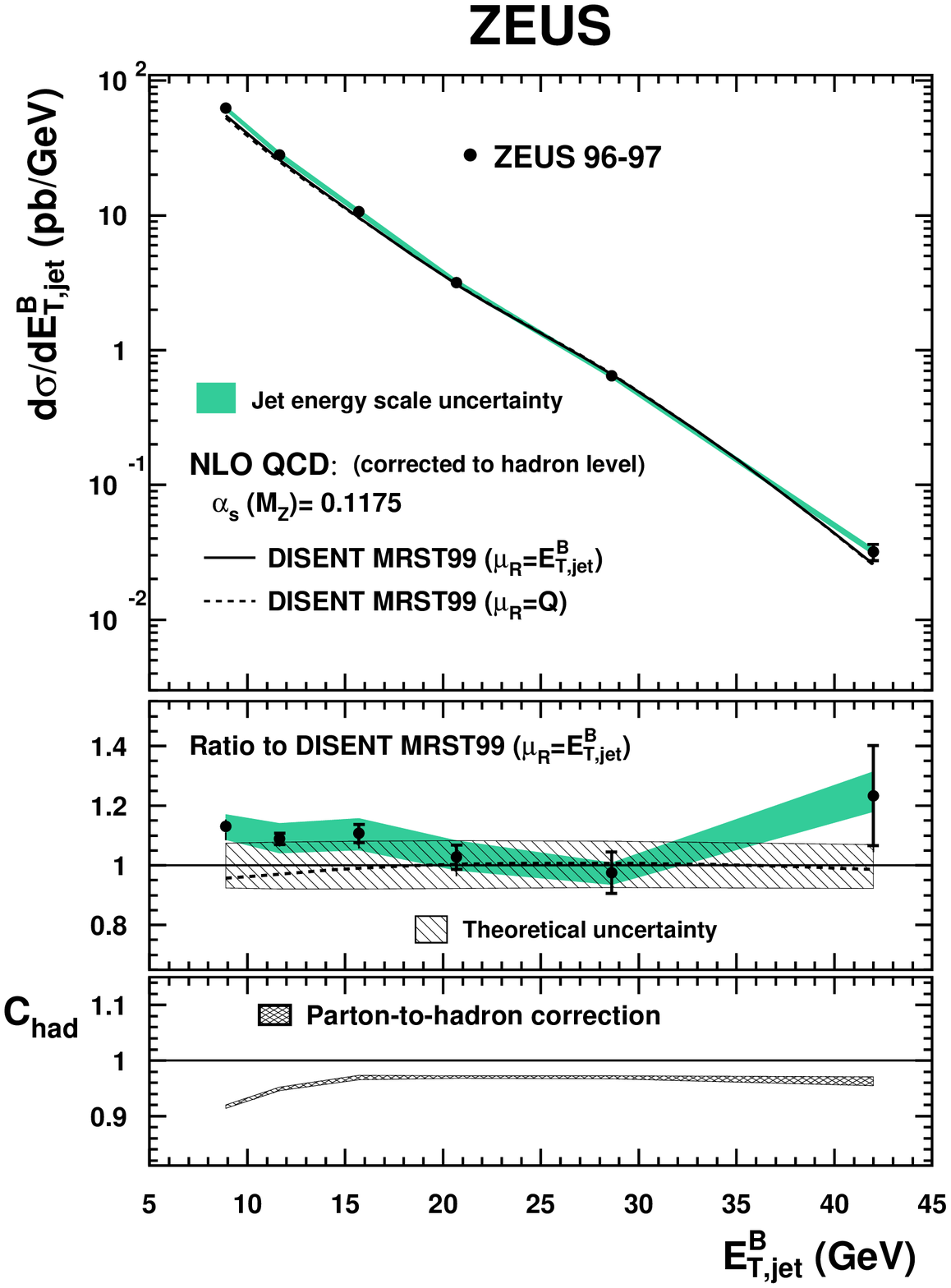,width=9.5cm}}
\put (4.2,0.0){\bf\small (a)}
\put (13.0,0.0){\bf\small (b)}
\end{picture}
\caption{Inclusive jet cross sections~\protect\cite{inczeus} as a
  function of (a) $\q2$ and (b) $\etjb$. 
  \label{one}}
\end{figure}

Figure~\ref{one} shows the inclusive jet cross section as a function
of $\q2$ and $\etjb$. The dots are the data and the error bars
represent the statistical and systematic uncertainties; the shaded band
displays the uncertainty on the absolute energy scale of the
jets. The measured cross sections have a steep fall-off, by five (four) 
orders of magnitude within the measured $\q2\ (\etjb)$ range. The
lines are the NLO calculations using DISENT with different choices of the
renormalisation scale ($\mu_R=Q$ or $\etjb$). The calculations
describe reasonably well the $\q2$ and $\etjb$ dependence of the cross
section for $\q2>500$ \g2\ and $\etjb>15$ GeV. At low $\q2$ and low
$\etjb$, the measurements are above the calculations by about $10\%$,
which is of the same size as the theoretical uncertainties (see
below). Therefore, for the extraction of $\as$, the phase space 
was restricted to high $\q2$ and high $\etjb$.

The experimental uncorrelated uncertainties for these cross sections
are small, $\sim 5\%$. The uncertainty coming from the absolute energy
scale of the jets is also small, $\sim 5\%$. The theoretical
uncertainties comprise $5\%$ from the absent higher orders, $3\%$ from the 
uncertainties of the proton PDFs and $5\%$ from the uncertainty in the
value of $\asz$ assumed. The parton-to-hadron corrections are $10\%$
with an uncertainty of $1\%$.

\subsection{Determination of $\as$}
The method~\cite{inczeus} used by ZEUS to extract $\as$ exploits the
dependence of the NLO calculations on $\asz$ through the matrix
elements ($\hat\sigma\sim A\cdot\as +B\cdot\as^2$) and the
proton PDFs ($\asz$ value assumed in the evolution). To take into account
properly this correlation, NLO calculations were performed using
various sets of PDFs which assumed different values of $\asz$. The
calculations were then parametrised as a function of $\asz$ in each
measured $\q2$ or $\etjb$ region. From the measured value
of the cross section as a function of $\q2$ in each region of $\q2$, a
value of $\asz$ and its uncertainty were extracted using the
parametrisations of the NLO calculations.

\begin{figure}[h]
\setlength{\unitlength}{1.0cm}
\begin{picture} (10.0,8.2)
\put (-0.3,0.0){\epsfig{figure=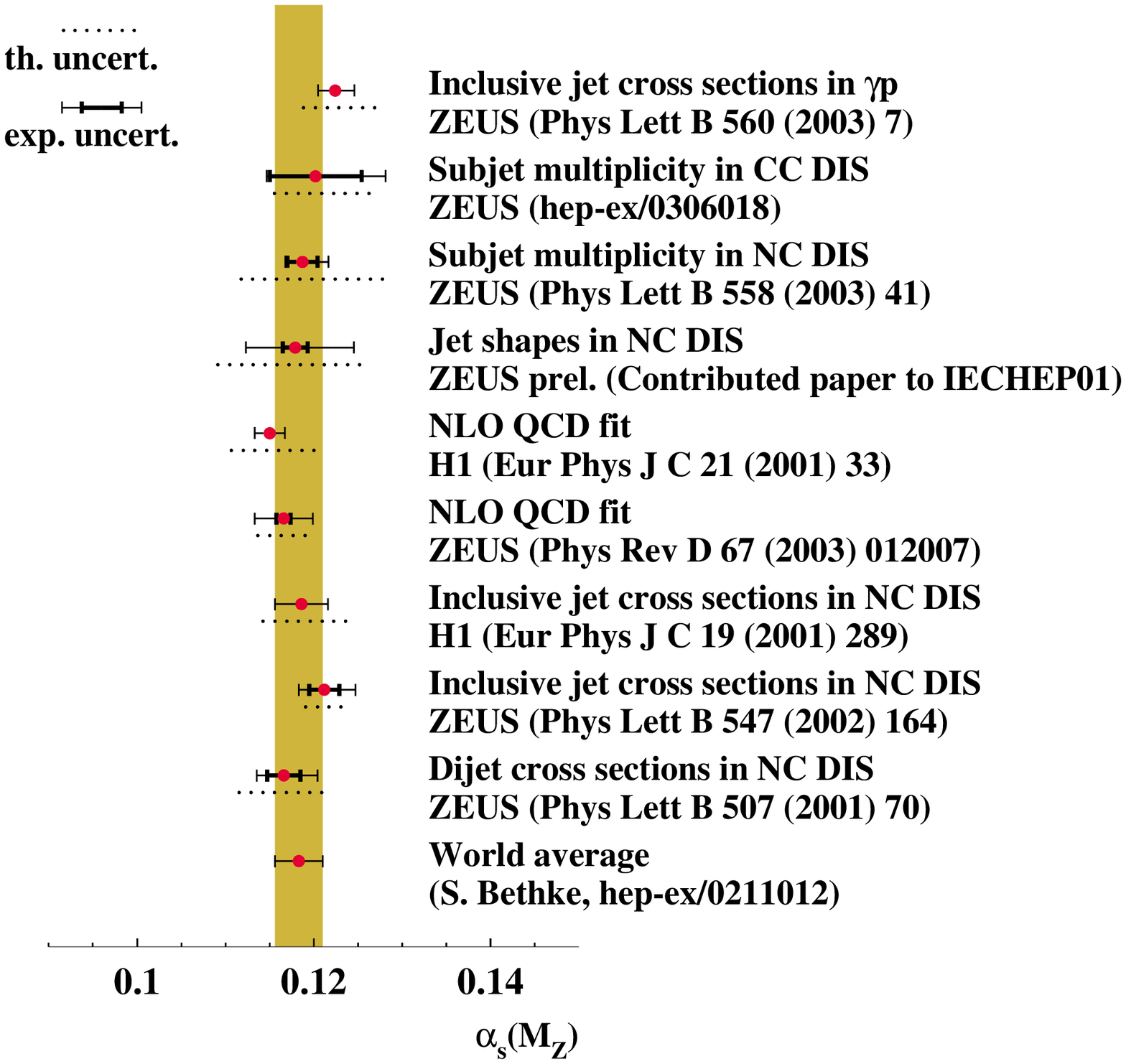,width=9.0cm}}
\put (8.7,-0.6){\epsfig{figure=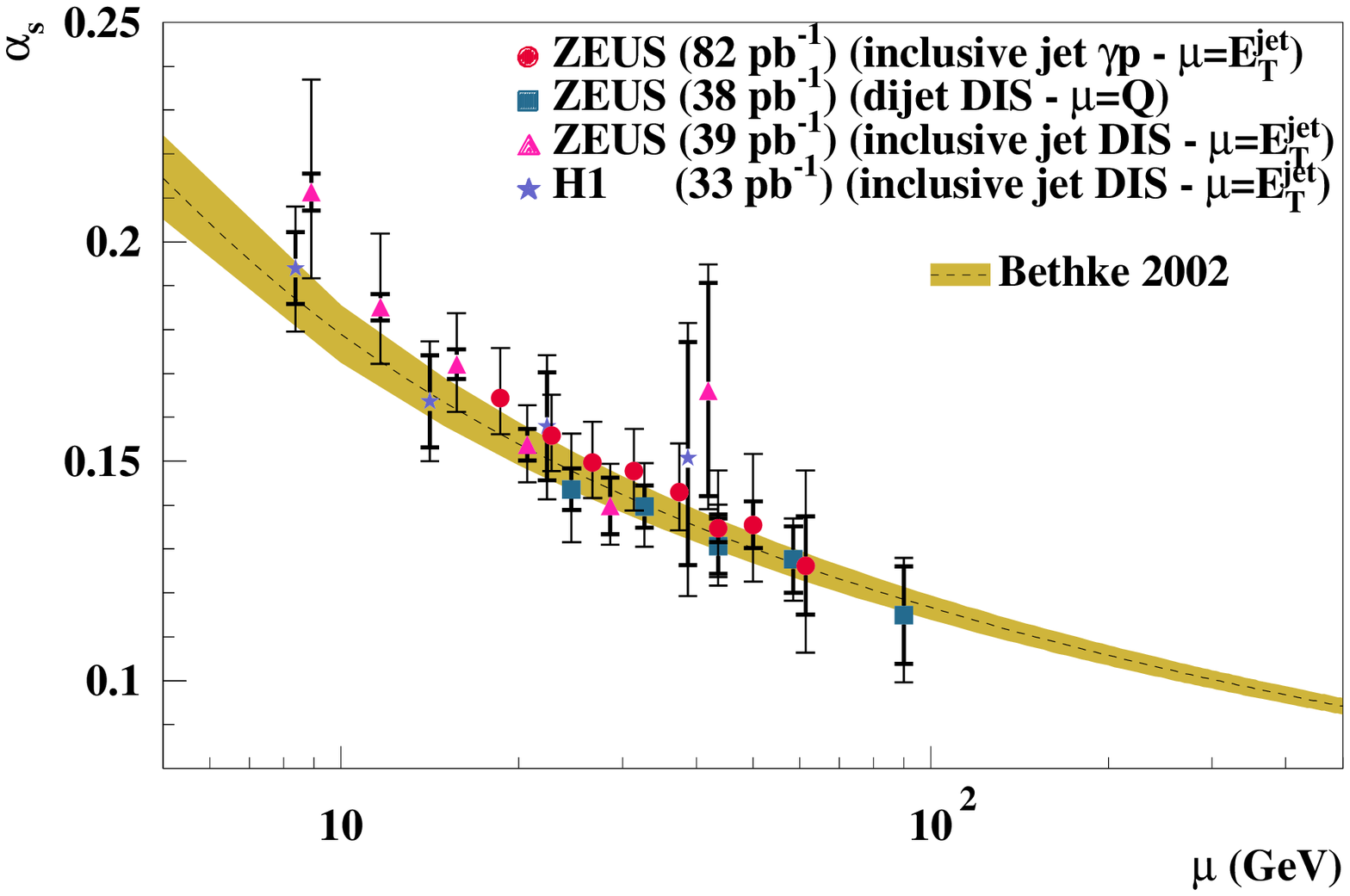,width=9.0cm}}
\put (1.8,0.0){\bf\small (a)}
\put (12.7,0.0){\bf\small (b)}
\end{picture}
\caption{(a) Summary of extracted $\asz$ values at HERA. (b) Summary of
  extracted $\as(\mu)$ values as a function of $\mu$ at HERA. 
  \label{two}}
\end{figure}

From the inclusive jet cross section for $\q2>500$ \g2, the value
$$\asmz{0.1212}{0.0017}{0.0031}{0.0023}{0.0027}{0.0028}$$
was extracted using the method explained above. The experimental
uncertainties are dominated by the uncertainty on the absolute energy
scale of the jets ($1\%$). The theoretical uncertainties are: $3\%$
from the absent higher orders, $1\%$ from the PDFs and $0.2\%$ from
the hadronisation corrections. This determination is compatible with
other independent extractions performed at HERA and with the current
world average (see figure~\ref{two}a). Further precision in the
extraction of $\asz$ from inclusive jet cross sections depends upon
further experimental and theoretical improvements.

The QCD prediction for the energy-scale dependence of $\as$ has been
tested by determining $\as$ from the measured differential cross
sections at different scales~\cite{inczeus}. From the measured cross
section as a function of $\etjb$, in each region of $\etjb$, a value
of $\as(\etjb)$ was extracted. The result, shown in figure~\ref{two}b
(triangles), is compatible with the running of $\as$ as predicted by
QCD (shaded band) over a large range in the scale. Figure~\ref{two}b
also shows other studies of the energy-scale dependence of $\as$ from
HERA: all the results are compatible with each other and with the QCD
prediction. This constitutes a test of the scale dependence of $\as$
between $\mu=8.4$ and $90$ GeV.

\section{Parton evolution at low $x$}
Dijet data in DIS may be used to gain insight into the parton dynamics at
low $x$. The evolution of the PDFs with the factorisation scale can be
described by the DGLAP evolution equations which sum the leading
powers of terms like $\as\log\q2$ in the region of strongly ordered
transverse momenta $\kt$. This presciption describes successfully jet
production at high $\q2$. However, the DGLAP approximation is expected
to break down at low $x$ since when $\log\q2\ll\log 1/x$, the terms
proportional to $\as\log 1/x$ become important and need to be
summed. This is done in the BFKL evolution equations; the integration
is taken over the full $\kt$ phase space of the gluons with no
$\kt$-ordering.

Another approach, the CCFM evolution equations with angular-ordered
parton emission, is equivalent to the BFKL approach for 
$x\rightarrow 0$ and reproduces the DGLAP evolution equations at large
$x$. Thus, the properties of the dijet system, which depend on the
dynamics of the ladder, can be studied to determine whether the
cascade has a $\kt$-ordered or unordered evolution.

Deviations from the DGLAP approach at small $x$ can be tested
experimentally. At small $x$ it is expected that parton emission along
the exchanged gluon ladder should increase with decreasing $x$. A
clear experimental signature of this effect would be that the two
outgoing hard partons are no longer back-to-back and so an excess
of events at small azimuthal separations should be observed.

Values of the azimuthal separation of the two hard jets in the
$\gamma^*p$ centre-of-mass frame, $\Delta\phi^*$, different than $\pi$
can occur in the DGLAP approach only when higher order contributions
are included. On the other hand, in the BFKL and CCFM approaches, the
number of events with $\Delta\phi^*<\pi$ should increase due to
the partons entering the hard process with large $\kt$.

\subsection{Azimuthal jet separation}
To test the predictions of the different approaches, dijet cross sections
have been measured~\cite{dijh1} using the $\kt$-cluster algorithm in
the longitudinally inclusive mode in the $\gamma^* p$ centre-of-mass
frame. The measurements were made in the kinematic region given by
$5<\q2<100$ \g2\ and $10^{-4}<x<10^{-2}$. The cross
sections refer to jets of $E_T^*>5$ GeV, $-1<\etalab<2.5$ and
$E_{T,{\rm max}}^*>7$ GeV, where $E_T^*$ is the jet transverse energy
in the $\gamma^* p$ centre-of-mass frame.

\begin{figure}[h]
\setlength{\unitlength}{1.0cm}
\begin{picture} (10.0,9.6)
\put (0.0,0.5){\epsfig{figure=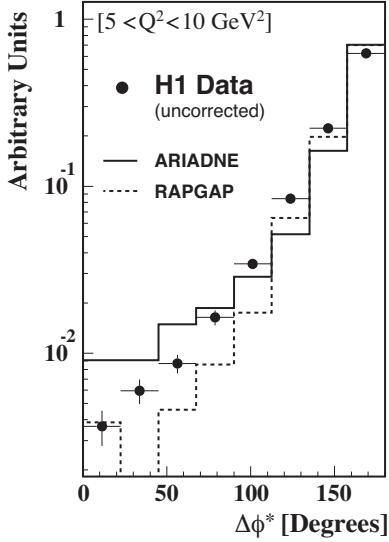,width=7.5cm}}
\put (8.0,0.1){\epsfig{figure=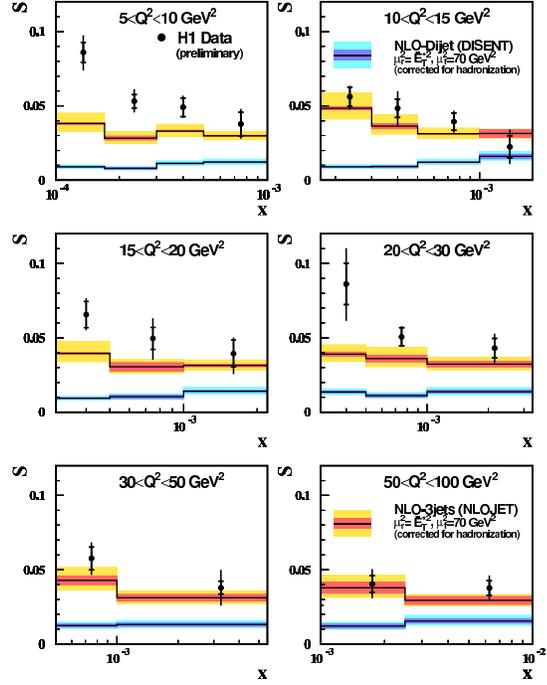,width=7.5cm}}
\put (3.5,0.0){\bf\small (a)}
\put (11.7,0.0){\bf\small (b)}
\end{picture}
\caption{(a) Azimuthal separation between
  jets~\protect\cite{dijh1}. (b) Ratio $S$ as a function of Bjorken $x$
  and $\q2$ compared with predictions from NLO QCD
  calculations~\protect\cite{dijh1}.
  \label{three}}
\end{figure}

Figure~\ref{three}a shows the measured dijet distribution as a
function of $\Delta\phi^*$. A significant fraction of events 
is observed at a small azimuthal separation. Since a measurement of a
multi-differential cross section as a function of $x$, $\q2$ and
$\Delta\phi^*$ would be very difficult due to large migrations, the
fraction of the number of dijet events with an azimuthal separation
between $0$ and $\alpha$, where $\alpha$ was taken as 
$\alpha=\frac{2}{3}\pi$, was measured instead. The fraction $S$,
defined as
$$S=\frac{\int_0^{\alpha} N_{\rm 2jet}(\Delta\phi^*,x,\q2)d\Delta\phi^*}{\int_0^{\pi} N_{\rm 2jet}(\Delta\phi^*,x,\q2)d\Delta\phi^*},$$
is better suited to test small-$x$ effects than a triple differential
cross section.

The measured fraction $S$ as a function of Bjorken $x$ in different
regions of $\q2$ is presented in figure~\ref{three}b. The data rise
towards low $x$ values, especially at low $\q2$. The NLO predictions
from DISENT, which contain $\kt$ effects only in the first order
corrections, are several standard deviations below the data and show
no dependence with $x$. On the other hand, the predictions of NLOJET,
which contain $\kt$ effects at next-to-lowest order, provide an accurate
description of the data at large $\q2$ and large $x$. However, they
fail to describe the increase of the data towards low $x$ values,
especially at low $\q2$.

\begin{figure}[h]
\setlength{\unitlength}{1.0cm}
\begin{picture} (10.0,10.5)
\put (-0.4,0.2){\epsfig{figure=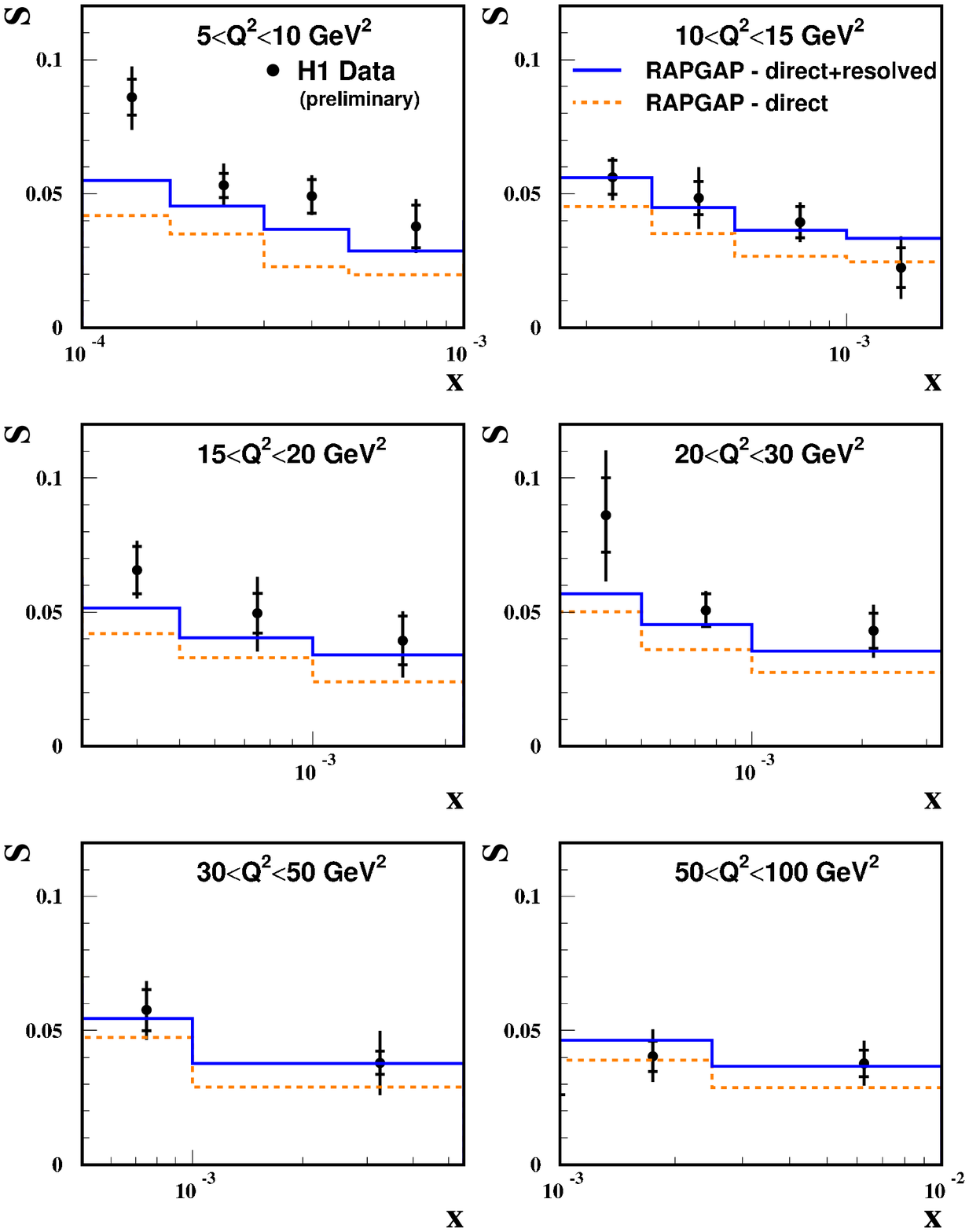,width=8.0cm}}
\put (8.5,0.2){\epsfig{figure=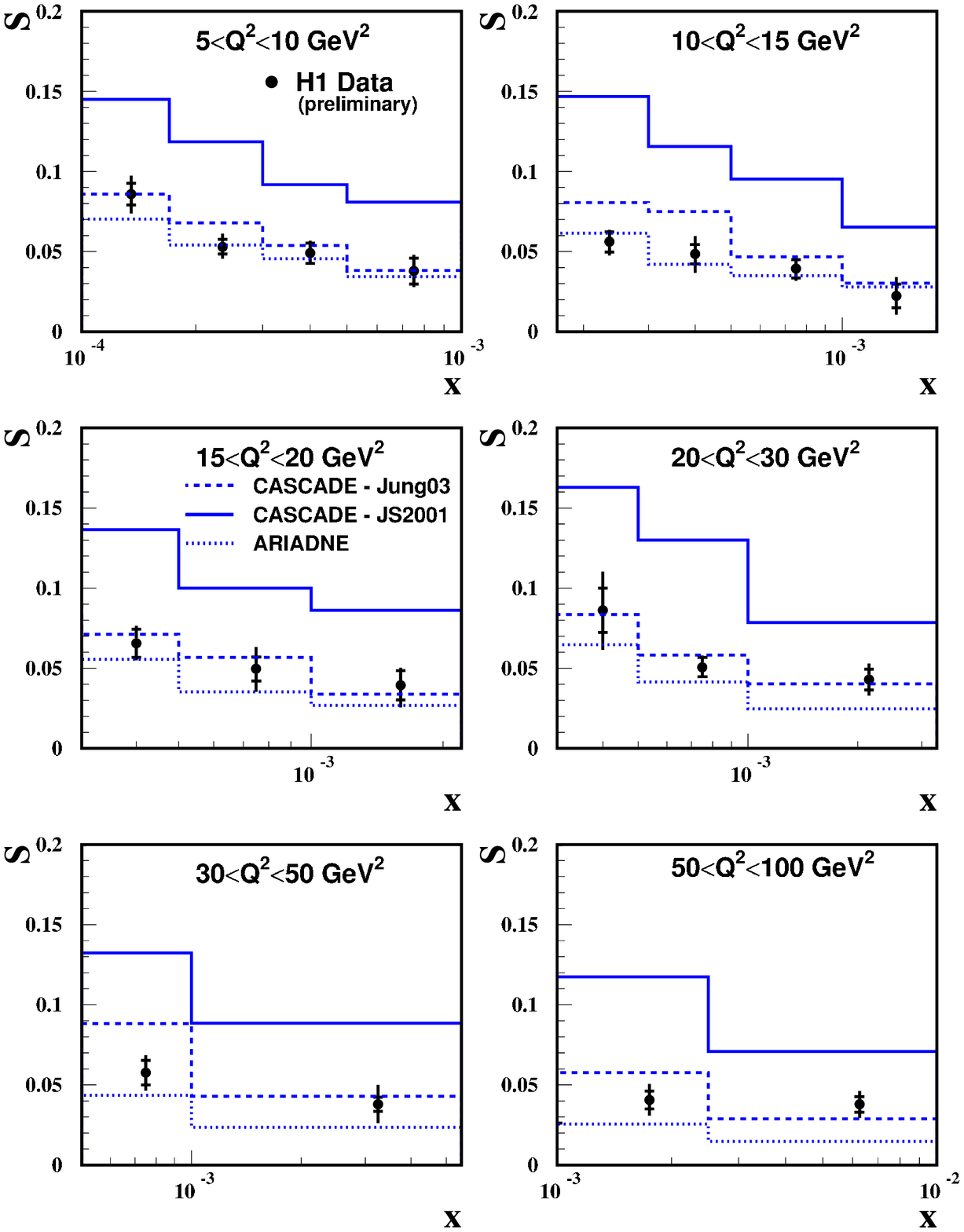,width=8.0cm}}
\put (3.7,0.0){\bf\small (a)}
\put (12.7,0.0){\bf\small (b)}
\end{picture}
\caption{Ratio~\protect\cite{dijh1} $S$ as a function of $x$ and $\q2$
  compared with predictions from RAPGAP (a) and ARIADNE and CASCADE (b).
  \label{four}}
\end{figure}

Figure~\ref{four}a shows the data compared with the predictions of
RAPGAP with direct only and resolved plus direct processes. A good
description of the data is obtained at large $\q2$ and large
$x$. However, there is a failure to describe the strong rise of the
data towards low $x$, especially at low $\q2$, even when including a
possible contribution from resolved virtual photon processes,
though the description in other regions is improved.

If the observed discrepancies are due to the influence of non-ordered
parton emissions, models based on the color dipole or the CCFM
evolution could provide a better description of the
data. Figure~\ref{four}b shows the data compared with ARIADNE (dotted
lines) and two predictions of CASCADE which use different sets of
unintegrated parton distributions. These sets differ in the way the
small-$\kt$ region is treated: in Jung2003 the full splitting
function, i.e. including the non-singular term, is used in contrast to
JS2001, for which only the singular term was considered. The
predictions of ARIADNE give a good description of the data at low $x$
and $\q2$, but fail to describe the data at high $\q2$. The predictions
of CASCADE using JS2001 lie significantly above the data in all $x$
and $\q2$ regions, whereas those using Jung2003 are closer to the
data. Therefore, the measurement of the fraction $S$ is sensitive to
the details of the unintegrated parton distributions.

\section{Internal structure of jets}
The investigation of the internal structure of jets gives insight into
the transition between a parton produced in a hard process and the
experimentally observable jet of hadrons. The internal structure of a
jet depends mainly on the type of primary parton from which it
originated and to a lesser extent on the particular hard scattering
process. QCD predicts that at sufficiently high $\etjet$, where
fragmentation effects become negligible, the jet structure is driven
by gluon emission off the primary parton and is then calculable in
pQCD. The lowest non-trivial order contribution to the jet
substructure is given by order $\alpha\as$ calculations.

The internal structure of the jets can be studied by means of
the mean subjet multiplicity. Subjets are resolved within a jet by
reapplying the $\kt$ algorithm on all particles belonging to the jet
until for every pair of particles the quantity 
$d_{ij}={\rm min}(E_{T,i},E_{T,j})^2\cdot((\eta_i-\eta_j)^2+(\varphi_i-\varphi_j)^2)$ 
is above $d_{\rm cut}=\yc\cdot(\etjet)^2$. All remaining clusters are
called subjets. The subjet structure depends upon the value chosen for the
resolution parameter $\yc$.

\begin{figure}[h]
\setlength{\unitlength}{1.0cm}
\begin{picture} (10.0,10.7)
\put (-0.2,-0.3){\epsfig{figure=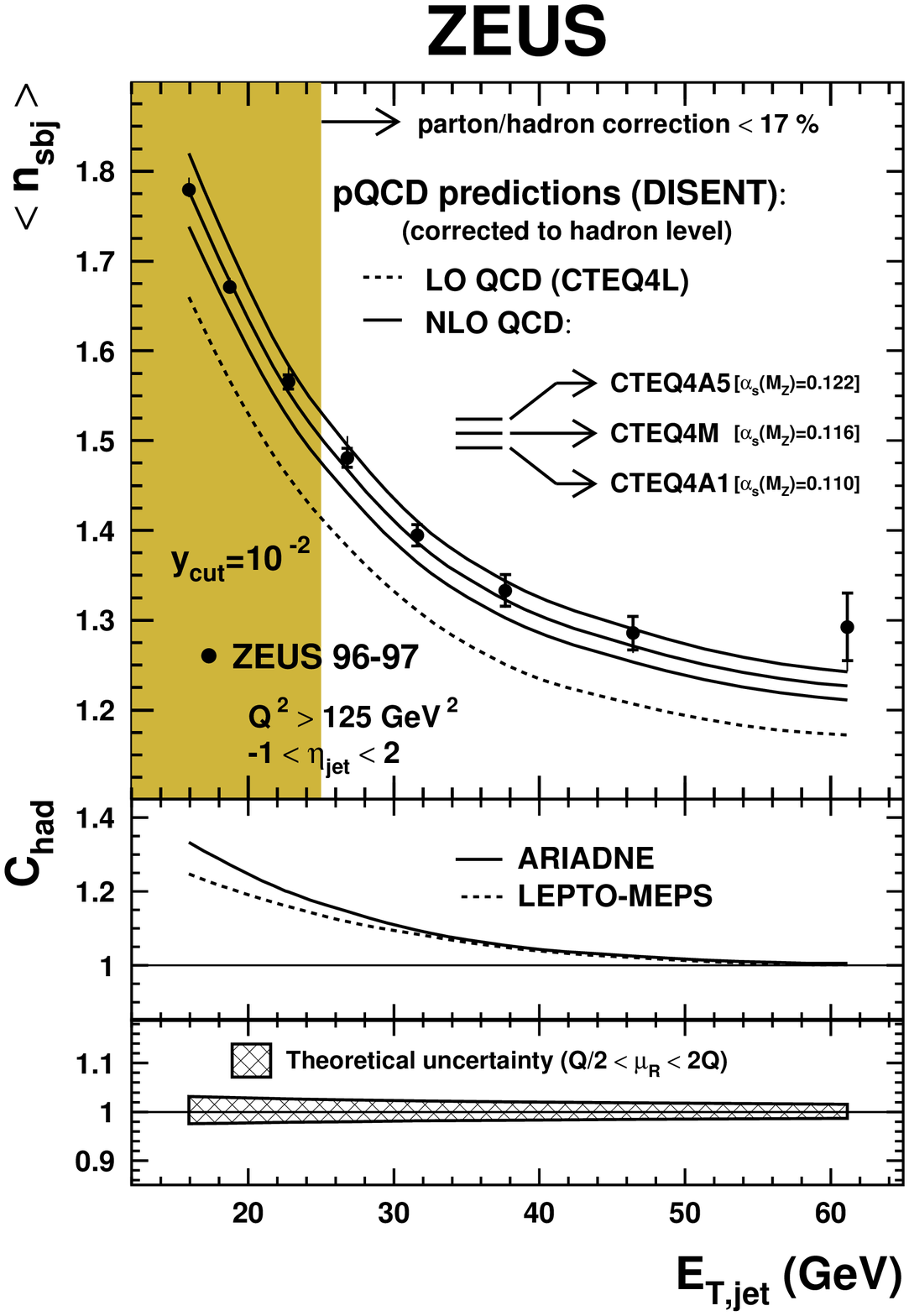,width=9.0cm}}
\put (8.4,0.2){\epsfig{figure=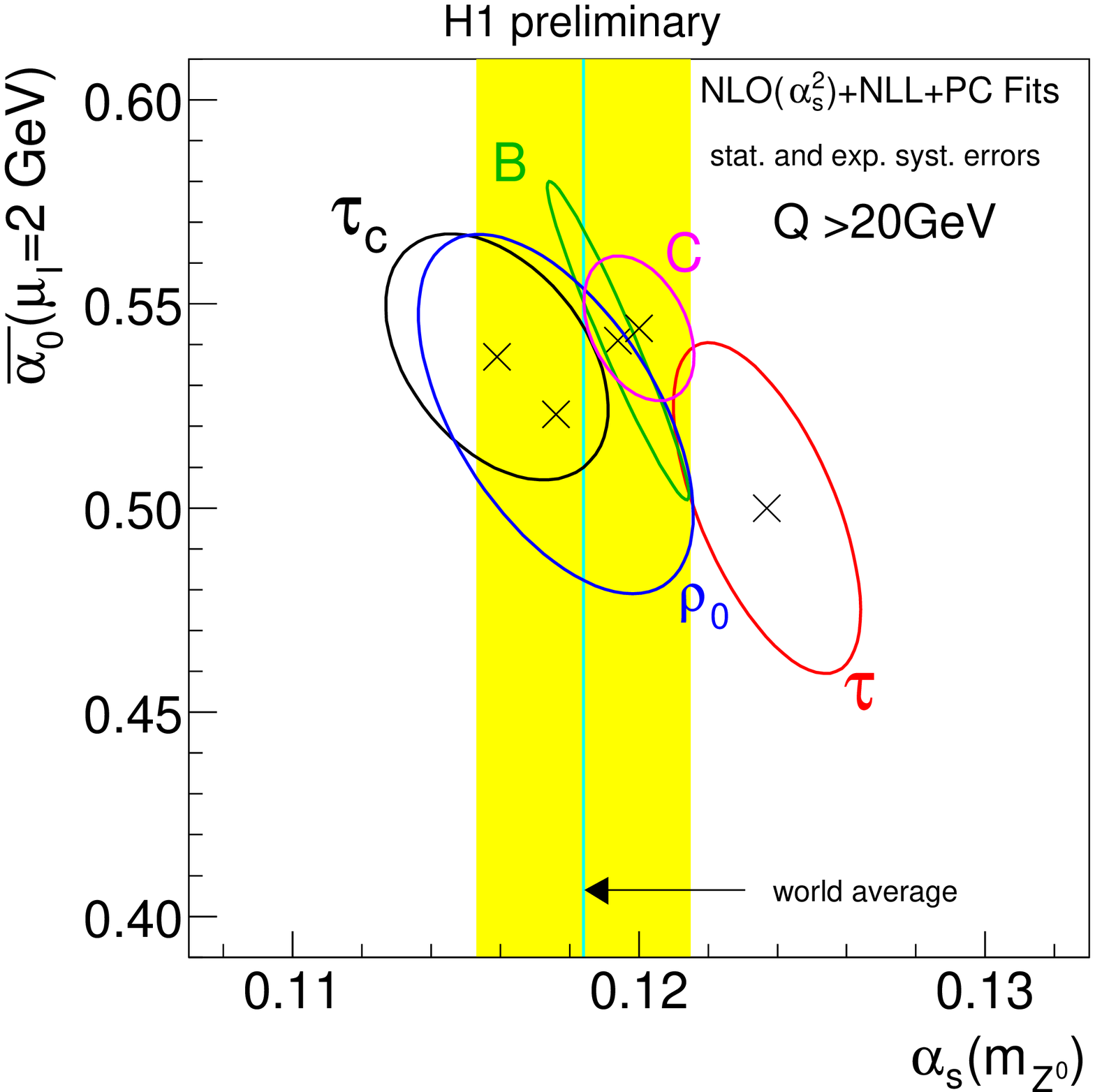,width=9.0cm}}
\put (4.2,0.0){\bf\small (a)}
\put (13.2,0.0){\bf\small (b)}
\end{picture}
\caption{(a) Mean subjet multiplicity~\protect\cite{inczeus} as a
  function of $\etjet$. (b) $1-\sigma$ contours in the
  $(\as,\bar\alpha_0)$ plane~\protect\cite{evh1}.
  \label{five}}
\end{figure}

The mean subjet multiplicity has been measured~\cite{subzeus} for jets
using the $\kt$ algorithm in the HERA laboratory frame with $\etjet$
above 15 GeV and $-1<\etajet<2$, in the kinematic range given by
$\q2>125$ \g2. Figure~\ref{five}a shows the mean subjet multiplicity
for a fixed value of $\yc$ of $10^{-2}$ as a function of $\etjet$. It
decreases as $\etjet$ increases, i.e. the jets become more
collimated. The experimental uncertainties are small (fragmentation
model uncertainty $<1\%$, the uncertainty on the absolute energy scale of
the jets is negligible). The detector and hadronisation corrections
are $<10\%$ and $<17\%$, respectively, for $\etjet>25$ GeV. The data are
compared to the LO and NLO predictions of DISENT. The LO calculation
fails to describe the data, whereas the NLO calculations provide a good
description. These measurements are sensitive to $\as$ and have been
used to extract a value of $\asz$. The result is
$$\asmz{0.1187}{0.0017}{0.0009}{0.0024}{0.0076}{0.0093}.$$
This value is compatible with the world average and with previous
measurements (see figure~\ref{two}a).

\section{Event shapes}
A complementary extraction of $\as$, also using the details of the
hadronic final state in DIS, comes from the study of the event shape
variables, like thrust or jet broadening. Event shape variables are
particularly sensitive to the details of the non-perturbative effects
of hadronisation and can be used to test the models for these
effects. Recently, new developments with regard to power-law
corrections have prompted a revived interest in the understanding of
hadronisation from first principles. In this type of analysis, the
data are compared to model predictions which combine NLO calculations
and the theoretical expectations of the power corrections, which are
characterised by an effective coupling $\bar\alpha_0$. Previous
results supported the concept of power corrections in the approach of
Dokshitzer \etal\ but a large spread of the results suggested that
higher order corrections were needed. Now, resummed NLL calculations
matched to NLO are available and so it is possible to study event
shape distributions instead of only their mean values.

Event shape distributions (thrust, jet broadening, the jet mass $\rho$
and the C parameter) have been measured~\cite{evh1} for particles in
the current hemisphere in the kinematic region given by $14<Q<200$ GeV
and $0.1<y<0.7$, where $y$ is the inelasticity variable. Predictions
consisting of NLO calculations using DISASTER++, resummed
calculations matched to NLO and power corrections have been fitted to
the data, leaving $\as$ and $\bar\alpha_0$ as free parameters. A good
description of the data by the predictions was obtained at high $\q2$,
though the description at low $\q2$ was poorer. Figure~\ref{five}b
shows the $1\sigma$-contour results from the fit in the
$\as$-$\bar\alpha_0$ plane. The spread observed in previous studies is
much reduced when the resummed calculations are included. A clear
anti-correlation between $\as$ and $\bar\alpha_0$ is found for all
variables. A universal value for $\bar\alpha_0$ of $0.5$ at the $10\%$
level was obtained, in agreement with the previous results, but with a
smaller spread. There is still a sizeable theoretical uncertainty for
both $\as$ and $\bar\alpha_0$, of the order of $5\%$, which is as
large as the experimental uncertainties. This uncertainty comes from
the absent higher order corrections.

\section{Conclusions}
HERA has become a unique QCD-testing machine due to the
fact that at large scales considerable progress in understanding
and reducing the experimental and theoretical uncertainties has led to
very precise measurements of the fundamental parameter of the
theory, the strong coupling constant $\as$. The use of observables
resulting from jet algorithms leads now to determinations that are as
precise as those coming from more inclusive measurements, such as
from $\tau$ decays. To obtain even better accuracy in the
determination of QCD, further improvements in the QCD calculations are
needed, e.g. next-to-next-to-leading-order corrections.

At low values of $x$ and $\q2$, considerable progress has also been 
obtained in understanding the mechanisms of parton emission, though
the interplay between the DGLAP, BFKL and CCFM evolution schemes has
still to be fully worked out. Further progress in this respect needs
both more experimental and more theoretical work.

\section*{Acknowledgments}
I would like to thank the organisers for providing a warm atmosphere
conducive to many physics discussions and a well organised
conference. Special thanks to my colleagues from H1 and ZEUS for their
help in preparing this report.

\end{document}